\documentclass[12pt]{iopart}

\usepackage{iopams}
\usepackage[]{graphicx}

\newcommand{\Exp}[1]{\mathbb{E}\left[#1\right]}

\begin{document}

\title[A statistical mechanics approach to de-biasing and uncertainty estimation in LASSO for random measurements]
{A statistical mechanics approach to de-biasing and uncertainty estimation in LASSO for random measurements}

\author{Takashi Takahashi and Yoshiyuki Kabashima}
\ead{takahashi.t.cc@m.titech.ac.jp}
\address{Department of Mathematical and Computing Science \\
Tokyo Institute of Technology\\
2-12-1, Ookayama, Meguro-ku, Tokyo, Japan}

\begin{abstract}
In high-dimensional statistical inference in which the number of parameters to be estimated is larger than that of the holding data, regularized linear estimation techniques are widely used. 
These techniques have, however, some drawbacks. 
First, estimators are biased in the sense that their absolute values are shrunk toward zero because of the regularization effect. 
Second, their statistical properties are difficult to characterize as they are given as numerical solutions to certain optimization problems.
In this manuscript, we tackle such problems concerning LASSO, which is a widely used method for sparse linear estimation, when the measurement matrix is regarded as a sample from a rotationally invariant ensemble.
We develop a new computationally feasible scheme to construct a de-biased estimator with a confidence interval and conduct hypothesis testing for the null hypothesis that a certain parameter vanishes.
It is numerically confirmed that the proposed method successfully de-biases the LASSO estimator and constructs confidence intervals and $p$-values by experiments for noisy linear measurements.

\end{abstract}

\maketitle
\tableofcontents

\section{Introduction}  
Estimating high-dimensional unknown variables from a limited number of data precisely and reliably is an important task in statistics, machine learning, signal processing, and so on.
    For instance, such demands arise in compressed sensing \cite{D2006,CT2006} and genomics \cite{Petal2010}.  
	 Since, in these problems, the number of parameters often far surpasses that of observed data, it is clear that some sparsity assumptions on the parameters are necessary to reasonably estimate them.
     Therefore, one needs to simultaneously solve two problems: variable selection, which seeks relevant (or non-zero) parameters for the data generation process, and parameter estimation.
     In the past few decades, a number of methods have been developed to tackle such problems.
     One of the most successful approaches is the least absolute shrinkage and selection operator (LASSO) \cite{T1996} method for high-dimensional linear regression problems in which the estimator is obtained by minimizing the $L_1$ norm regularized likelihood function.
     As LASSO estimators can be easily obtained by versatile algorithms for convex optimization \cite{CT2006, CT2005} and have appealing consistency properties \cite{BRT2009,DMM2009,KWT2009}, they have received considerable attention.
     
			Specifically, let us consider the linear measurement model:
            \begin{equation}
            	y_i = \bi{a}_i^\top \bi{x}_0 + \xi_i, \,\xi_i\sim_{\mathrm{i.i.d}}\mathcal{N}\left(0, \sigma^2\right), \, i = 1,2,...,M,
                \label{eq:linear_regression}
            \end{equation}
         where $\bi{x}_0\in \mathbb{R}^N$ and $\bi{a}_i\in \mathbb{R}^N$ are the parameter (signal) and measurement vectors, respectively, $\sigma^2 \in \mathbb{R}$ is a parameter that describes the strength of the measurement noise, and $\mathcal{N}(\mu, \sigma^2)$ is the normal distribution with mean $\mu$ and variance $\sigma^2$.
         Notation $\top$ means the operation of matrix/vector transpose. 
         In matrix notation, this model is expressed as 
         \begin{equation}
         	\bi{y} = A\bi{x}_0 + \bi{\xi}, \label{eq:linear_regression_matrix_form}
         \end{equation}
         where $\bi{a}_i^\top$ corresponds to the $i$'-th row of the matrix $A\in \mathbb{R}^{M\times N}$.
         $A$ is called the observation or measurement matrix by cases.
         The objective of high-dimensional linear regression is to find the parameter vector $\bi{x}_0$, where the number of measurements $M$ is smaller than that of the parameter $N$.
         Note that in this high-dimensional setting, one cannot obtain a true solution with simple linear algebra because $A^\top A$ is not invertible; by contrast, in the classical setting where $M > N$, the unique unbiased estimator is easily obtained as $\widehat{\bi{x}}_{\mathrm{classical}} = (A^\top A)^{-1}A^\top \bi{y}$ by using the least squares method.
     		To achieve this aim, LASSO seeks an estimator by solving an optimization problem that imposes sparsity via an $L_1$ penalty:　
            \begin{equation}\label{eq:LASSO}
            	\widehat{\bi{x}}^{\mathrm{LASSO}}(\bi{y}, A;\lambda) \equiv \arg\min_{\bi{x}}\left[\frac{1}{2}\|\bi{y} - A\bi{x} \|_2^2 + \lambda \| \bi{x}\|_1\right],
            \end{equation}
            where $\lambda$ is a hyperparameter that controls the strength of the regularization.
            This convex optimization problem can be solved efficiently by using various versatile algorithms. 
            Although LASSO might be seen as simple heuristics, it has an appealing consistency property: in a certain sparsity condition on the true parameter $\bi{x}_0$ and an appropriate control of the regularization strength $\lambda$, the LASSO solution and $\bi{x}_0$ are consistent in the sense that $\|\widehat{\bi{x}}^{\rm LASSO} - \bi{x}_0 \|_2^2/N$ vanishes as the measurement ratio $\gamma\equiv M/N$ tends to infinity.
            For a more comprehensive review of LASSO in the context of high-dimensional settings, see \cite{BS2011}.

			Unfortunately, LASSO also has some drawbacks.
        First, the LASSO solution is biased as long as $\lambda>0$ is finite.
        The amplitude of the LASSO estimator $\widehat{\bi{x}}^{\rm LASSO}$ is shrunk toward zero by the regularization term and its absolute value is typically smaller than that of the true parameter $\bi{x}_0$ even in an ideal sparsity assumption. 
			Second, no explicit form of the distribution is available for the estimator, as it is just expressed as a numerical solution of (\ref{eq:LASSO}). 
        Consequently, one can neither construct confidence intervals nor perform hypothesis testing for the null hypothesis that a certain element of the parameter vanishes. These bottlenecks are considered to be problematic in real applications in which the statistical reliability of the estimation result should be assessed.
        This situation is different from the one of classical statistics in which one can analytically obtain an unbiased estimator and its distribution. 
            
        To resolve the problems stated above, in this study, we develop a new scheme for de-biasing and uncertainty estimation in the LASSO estimation in the case that the observation matrix $A$ is generated from rotationally invariant random matrix ensembles, which are concretely defined in the next section. 
        The uncertainty addressed in this study concerns the randomness that arises from the random observation matrix $A$ and measurement noise $\bi{\xi}$.
        Our approach is based on a careful observation of the replica analysis of LASSO and an advanced mean-field method known as expectation consistent approximation or the adaptive Thouless--Anderson--Palmer (TAP) approach \cite{OW, OW2, OW3} developed in machine learning \cite{M} and statistical mechanics. 
        We numerically show that the proposed algorithm effectively de-biases the LASSO estimator and estimates its uncertainty.
                
  The rest of this manuscript is organized as follows.
  In section \ref{sec:num2}, we explain the problem setting.
  In section \ref{sec:num3}, we describe the result of the replica analysis of LASSO and its physical implications. 
  Then, the design of our scheme is introduced.
  The derivation of the free energy density is in \ref{app:num1}.
  In section \ref{sec:num4}, the proposed scheme is numerically tested by experiments for noisy linear measurements using various matrix ensembles.
  The last section provides a summary.
    
\section{Problem Setting} \label{sec:num2}

	\subsection{Model specification}
   In this study, we focus on random design models of (2), in which $A$ is a random matrix and the true parameter vector $\bi{x}_0$ is sparse in the sense that the number of its non-zero components is limited to $\varrho N$ ($0\leq \varrho < 1$). 
   More precisely for $A$, we assume that for eigenvalue decomposition $A^\top A = O D O^\top$, $O$ can be regarded as a random sample from the uniform distribution of the $N\times N$ orthogonal matrices and the empirical eigenvalue distribution $\sum_{i=1}^N \delta(\lambda - \lambda_i) /N $, where $\{\lambda_i\}_i$ are the eigenvalues of $A^\top A$, converges to a certain distribution $\rho(\lambda)$ in the limit $N\to\infty$ with probability one.
    
    \subsection{De-biasing and uncertainty estimation in LASSO}
    Let $\widehat{\bi{x}}^{\mathrm{LASSO}}(\bi{y}, A;\lambda)$ be the LASSO estimator for the given $\bi{y}, A$, and $\lambda$.
	We are interested in the two problems associated with $\widehat{\bi{x}}^{\mathrm{LASSO}}(\bi{y}, A; \lambda)$.
    The first problem is that the LASSO estimator is biased. 
    In other words, $\left|\mathbb{E}\left[\widehat{x}^{\mathrm{LASSO}}_i\right]_{A, \bi{\xi}}-x_{0,i}\right|$,$\,$ $(i=1,2,...,M)$ remains finite for $\lambda > 0$ because of the shrinkage effect caused by the regularization term $\lambda\|\bi{x}\|_1$.
    The second is that the LASSO estimator does not have an explicit form of the distribution. As a consequence, one can neither construct a confidence interval nor compute a $p$-value to conduct hypothesis testing for the null hypothesis that a certain parameter vanishes. 
    
	 In response to the aforementioned problems, we construct the following quantities.
    The first quantity is the de-biased estimators $\{\widehat{x}_i^{\rm debiased}\}_i$ that have confidence intervals $\{\mathcal{I}_i(\alpha_{\rm sig})\equiv[\widehat{x}_i^{\rm debiased} - L_i(\alpha_{\rm sig}), \widehat{x}_i^{\rm debiased} + U_i(\alpha_{\rm sig})]\}_i$ with significance $\alpha_{\rm sig}$.  
    The term \emph{de-biased} means that this estimator coincides with the true parameter on average: $\mathbb{E}[\widehat{x}_i^{\rm debiased}]_{A,\bi{\xi}} = x_{0,i}$. 
    The second quantity is the $p$-values to test whether the LASSO estimator is zero or not.
    We are interested in hypothesis testing with the null hypothesis $H_{0,i}: x_{0,i} = 0$.
    The confidence intervals concerning the de-biased estimators and hypothesis testing via $p$-values assess the uncertainty in LASSO.
	
	In the past few years, several researchers have been working on the issue closely related to that stated here \cite{JM2014,GBRD2014,ZZ2014}. 
   These studies discuss de-biasing and hypothesis testing in high-dimensional statistics for a fixed observation matrix where the randomness comes from the measurement noise, under tight sparsity assumptions on a true sparse signal, which corresponds to the $\varrho \to 0 $ limit in the current setting.
   In contrast to these studies, we concentrate on the case that the randomness comes from both the random observation matrix and the measurement noise without an explicit sparsity assumption on the true parameter keeping $\varrho \sim O(1)$.
   
\section{A Statistical Mechanics Approach} \label{sec:num3}

    \subsection{Replica analysis for general rotationally invariant random design matrices and its physical implications}
    To investigate how the LASSO solution depends on the true solution, observation matrix, and measurement noise, we first evaluate the free energy density corresponding to the LASSO Hamiltonian $H(\bi{x}) \equiv \|\bi{y} - A\bi{x}\|_2^2/2 + \lambda \|\bi{x}\|_1$ at a zero-temperature limit:
    \begin{equation} \label{eq:free_energy_density}
    	f(\lambda) \equiv -\lim_{\beta\to\infty}\lim_{N\to\infty}\frac{1}{N\beta}
        	\mathbb{E}\left[\ln Z(\bi{y},A;\lambda)\right]_{A,\bi{\xi}},
     \end{equation}
     where $\beta$ is the inverse temperature and $Z$ is the partition function:
     \begin{equation}
        Z(A, \bi{y};\lambda) = \int \exp\left(-\frac{\beta}{2}\|\bi{y} - A\bi{x}\|_2^2 - \beta\lambda\|\bi{x}\|_1\right)d\bi{x}. \label{eq:partition_function}
    \end{equation}
    We take the limit $N\to\infty$ with $\gamma=M/N\sim O(1)$ fixed.
    In the zero-temperature limit $\beta\to\infty$, the Boltzmann distribution $e^{-\beta H(\bi{x})}/Z$ is dominated by the configurations of the LASSO solution (\ref{eq:LASSO}).
    Hence, one can evaluate how the LASSO estimator depends on $\bi{x}_0, A, \bi{\xi}$ by analyzing the macroscopic behavior of the typical free energy density (\ref{eq:free_energy_density}) using statistical mechanics.
    
   Since the Hamiltonian defined above has a mean-field nature in the sense that all the variables are weakly connected, the free energy density (\ref{eq:free_energy_density}) can be evaluated by using the replica method:
   \begin{eqnarray}\label{eq:free_energy_density_saddle}
   \fl
   	f = \mathop{\rm extr}_{\chi,\widehat{\chi},Q,\widehat{Q}, m, \widehat{m}}\left[
     G'(-\chi;J)(Q-2m + \varrho -\chi \sigma^2) + \frac{\gamma}{2}\sigma^2
    - \frac{\widehat{Q}Q}{2}  + \frac{\widehat{\chi}\chi }{2} + \widehat{m}m  \right.\nonumber\\
    \left.+ \lim_{N\to\infty}\frac{1}{N}\sum_{i=1}^N \int\min_{x_i} \left\{-\frac{\widehat{Q}}{2}x_i^2 + \left(\widehat{m}x_{0,i} + \sqrt{\widehat{\chi}}z_i\right)x_i - \lambda \left|x_i\right| \right\}  Dz_i 
    \right],
   \end{eqnarray}
   where $\mathop{\rm extr}_{\chi,\widehat{\chi},Q,\widehat{Q}, m, \widehat{m}} \mathcal{F} (\chi,\widehat{\chi},Q,\widehat{Q}, m, \widehat{m})$ denotes the extremization of the function $\mathcal{F}$ with respect to its arguments and $G^{\prime}(x;J)$ is the derivative of $G(x;J)$ with respect to $x$.
   We have defined $\int (...) Dz, J, G(x)$ as follows:
   \begin{eqnarray}
	   \int (...) Dz\equiv\int (...) \frac{e^{-\frac{z^2}{2}}}{\sqrt{2\pi}}dz, \\
       J\equiv A^\top A, \\
       G(x;J) \equiv \mathop{\rm extr}_z \left[-\int \rho_J (s)\ln \left|z - s\right|ds  + \frac{zx}{2}\right] - \frac{1}{2}\ln x - \frac{1}{2}, \label{eq:def_G}
   \end{eqnarray}
   where $\rho_J(s)$ is the asymptotic eigenvalue distribution of $J$.
   The derivative of the function $G(x;J)$ has the following form:
   \begin{equation}\label{eq:def_G_prime}
   	G^\prime(x;J) = \frac{1}{2}\left(z(x) - \frac{1}{x}\right),
   \end{equation}
   where $z(x)$ is implicitly determined by the extremal condition of (\ref{eq:def_G}):
   \begin{equation}\label{eq:stieltjes_transform}
   	x = \mathcal{S}_J(z(x)) \equiv \int\frac{\rho_J(\lambda)}{z(x)-\lambda}d\lambda.
   \end{equation}
   The transformation $\mathcal{S}_J$ that appears in (\ref{eq:stieltjes_transform}) is called the Stieltjes transformation of $\rho_J$.
   The introduced function $G$ is connected to the R-transform $\mathcal{R}_J(\cdot)$ of the asymptotic eigenvalue distribution of $J$ in studies of free probability theory \cite{V1986}: $G(x;J)=\int_0^{x} \mathcal{R}_J(t) dt$.
   \ref{app:num1} provides a brief derivation of the free energy density (\ref{eq:free_energy_density_saddle}).
   
   The connection between the free energy density (\ref{eq:free_energy_density_saddle}) and macroscopic observables is as follows. 
   At the extremum, $Q, m$, and $\chi$ correspond to the macroscopic physical observables: $Q=\Exp{\langle |\bi{x}|^2\rangle}_{A,\bi{\xi}}/N$, $m=\Exp{\langle\bi{x}_0^\top \bi{x}\rangle}_{A,\bi{\xi}}/N$, and $\chi = \beta\Exp{\langle|\bi{x}|^2\rangle - \left|\langle\bi{x}\rangle\right|^2}_{A,\bi{\xi}}/N$. Each of these corresponds to the self-overlap, the overlap between the LASSO solutions and true solutions, and the macroscopic susceptibility. The notation $\langle ... \rangle$ represents the Boltzmann average in the zero-temperature limit: $\langle...\rangle\equiv \lim_{\beta\to\infty}\int(...)e^{-\beta H(\bi{x})}d\bi{x}/Z $. In addition, from direct calculation, one can show the following relationships between the free energy density, regularization term, and residual sum of squares:
   \begin{eqnarray}
   f &= \frac{\gamma}{2} \overline{\mathrm{RSS}} + \bar{r}, \label{eq:free_energy_density_rss_rbar}\\
   	\bar{r} &\equiv \Exp{\left\langle\frac{1}{N}\sum_{i=1}^N \left|x_i \right| \right\rangle}_{A,\bi{\xi}} = 
     \widehat{\chi}\chi + \widehat{m}m - \widehat{Q}Q, \label{eq:regularization_term}\\
     \overline{\mathrm{RSS}} &= \Exp{{\rm RSS}}_{A,\bi{\xi}}\equiv \Exp{\left\langle\frac{1}{M}\left\|\bi{y} - A\bi{x} \right\|_2^2\right\rangle}_{A,\bi{\xi}},\label{eq:rss_def}  \\
     &= \frac{2}{\gamma}\left[G'(-\chi; J ) (Q- 2m + \varrho - \chi\sigma^2) + \frac{\gamma}{2}\sigma^2 - \frac{1}{2}\chi\widehat{\chi}\right],\label{eq:rss}
   \end{eqnarray}
   where $\bar{r}$ and $\overline{\mathrm{RSS}}$ represent the per-element average of the regularization term and residual sum of squares, respectively.
   By using the relationships (\ref{eq:regularization_term}) and (\ref{eq:rss}) and the extremal condition concerning $\widehat{Q}, \widehat{m}, \widehat{\chi}$, the conjugate fields $\widehat{Q}, \widehat{m}, \widehat{\chi}$ can be represented via the macroscopic physical variables:
   \begin{eqnarray}
    \fl
    \widehat{\chi} = \frac{\gamma G''(-\chi; J)}{G'(-\chi; J) - G''(-\chi; J)\chi}\overline{\mathrm{RSS}}  + \frac{-G''(-\chi; J)\gamma + 2\left(G'(-\chi; J)\right)^2}{G'(-\chi; J) - G''(-\chi; J)\chi}\sigma^2,  \label{eq:chi_hat}\\
       \fl
   	\widehat{Q} = \widehat{m} = 2G'(-\chi;J).
   \end{eqnarray}
   Here, $\chi, G^{\prime}(-\chi;J)$ and $G^{\prime\prime}(-\chi;J)$ are given as follows:
   \begin{eqnarray}
   \chi = -\mathcal{S}_J(z(-\chi)) \label{eq:chi_stieltjes},\\
   	G^{\prime}(-\chi;J) = \frac{1}{2}\left(z(-\chi) + \frac{1}{\chi}\right), \label{eq:G_prime}\\
    G^{\prime\prime}(-\chi;J) = \frac{1}{2}\left(z^{\prime}(-\chi) + \frac{1}{\chi^2}\right),\label{eq:G_two_prime}
   \end{eqnarray}
   where $z^{\prime}(-\chi)$ is obtained from the derivative of equation (\ref{eq:stieltjes_transform}):
   \begin{equation}\label{eq:z_prime}
   	z^{\prime}(-\chi) = -\left[\int \frac{\rho_J(\lambda)}{(z(-\chi)-\lambda)^2}d\lambda\right]^{-1}.
   \end{equation}
   
\begin{figure}[t]
      \centerline{
      \includegraphics[width=16cm]{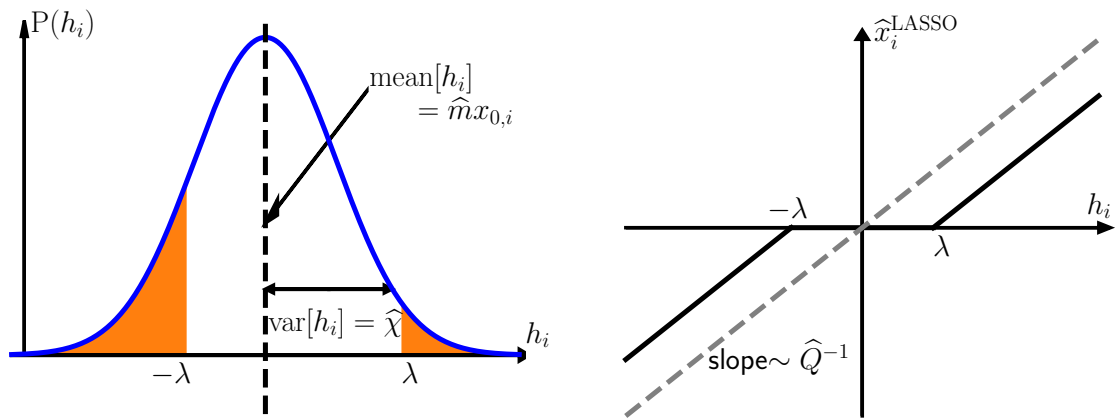}
      }
     \caption{Left: The distribution of the local fields.　The shaded region corresponds to the probability that the LASSO solution is active. Each local field is distributed according to the normal distribution $\mathcal{N}(\widehat{m}x_{0,i}, \widehat{\chi})$. In this example, $x_{0,i}<0$. Right: Local field dependence of the LASSO solution. The LASSO solution is determined by acting the soft-thresholding operator on the local field.}
     \label{fig:fig1}       
  \end{figure}

   The minimization problem in equation (\ref{eq:free_energy_density_saddle}) corresponds to the effective single body problem, which determines the value of the local magnetization $\langle x_i\rangle$. 
   Splitting into effective single body problems from the original multi-body estimation problem is called the \emph{decoupling principle} in the literature on information theory \cite{GV2006,RFG2009}.
   A comparison with the TAP/cavity analysis indicates that $h_i\equiv \widehat{m}x_{0,i} + \sqrt{\widehat{\chi}}z_i$ and $\widehat{m}$ correspond to the local field and Onsager reaction coefficient, respectively \cite{OS2001}. 
   Here, $z_i\sim\mathcal{N}(0,1)$ effectively represents the randomness that comes from the observation matrix and measurement noise.
   Figure \ref{fig:fig1} schematically shows the distribution of the local fields and how the local field determines the LASSO solution.
   Each local field is distributed according to the normal distribution $\mathcal{N}(\widehat{m}x_{0,i}, \widehat{\chi})$ and the LASSO solution is obtained by acting the soft-thresholding operator ${\rm ST}_{\lambda, \widehat{Q}}$ on it: 
   \begin{equation}
   	\widehat{x}_i^{\mathrm{LASSO}} = {\rm ST}_{\lambda, \widehat{Q}}(h_i)\equiv \frac{h_i - \lambda \mathrm{sgn}(h_i)}{\widehat{Q}}\Theta\left(\left|h_i \right| - \lambda\right),
   \end{equation}
   where $\Theta(z)$ is Heaviside's step function.
   
	The LASSO solution takes a non-zero value if the amplitude of the corresponding local field is larger than $\lambda$. Conversely, if and only if it is smaller than $\lambda$, the LASSO solution is exactly zero.
    Hereafter, we call the non-zero and zero components of the LASSO solution the \emph{active} and \emph{inactive} components, respectively.
    
    The above observations indicate that once the local fields and $\widehat{m}, \widehat{\chi}$ are estimated from the LASSO solutions, one can construct an intended $p$-value ${\rm P}_i$ as 
    \begin{eqnarray}
    	{\rm P}_i \equiv 2\left\{1- \Phi\left(\frac{h_i}{\sqrt{\widehat{\chi}}}\right)\right\} ,\label{eq:p-value}\\
       \Phi(x) \equiv \int_{-\infty}^x \frac{e^{-\frac{t^2}{2}}}{\sqrt{2\pi}}dt,
    \end{eqnarray}
    and an unbiased estimator as
    \begin{equation}\label{eq:debiased_estimator}
    	\widehat{x}_i^{\rm debiased} \equiv \frac{h_i}{\widehat{Q}},
    \end{equation}
    with a confidence interval 
    \begin{equation} \label{eq:confidence_interval}
	    \mathcal{I}_i(\alpha_{\rm sig}) = \left[\frac{h_i}{\widehat{Q}} - \Phi^{-1}\left(1-\frac{\alpha_{\rm sig}}{2}\right)\frac{\sqrt{\widehat{\chi}}}{\widehat{Q}}, \, \frac{h_i}{\widehat{Q}} + \Phi^{-1}\left(1-\frac{\alpha_{\rm sig} }{2}\right)\frac{\sqrt{\widehat{\chi}}}{\widehat{Q}}\,\right],
    \end{equation}
    of significance $\alpha_{\rm sig}$. 
    These are the key observations for the design of our scheme.

    \subsection{Adaptive TAP approach to constructing local fields and their variances from LASSO solutions}
    \subsubsection{Derivation of the adaptive TAP equations:}
    To derive the relation between the LASSO solution $\widehat{\bi{x}}^{\mathrm{LASSO}}(=\langle\bi{x}\rangle)$ and the local fields, let us consider Gibbs free energy:
  \begin{equation}\label{eq:Gibbs_raw}
    	G(\bi{m}) \equiv \mathop{\rm extr}_{\bi{h}}\left[
        	\bi{h}^\top \bi{m} - \frac{1}{\beta}\ln \left\{
            	e^{-\frac{\beta}{2}\|\bi{y} - A\bi{x}\|_2^2 + \beta\bi{h}^\top\bi{x} -\beta\lambda\|\bi{x}\|_1}d\bi{x}
            \right\}
        \right].
    \end{equation}
    The average $\langle\bi{x}\rangle$ is determined as the global minimizer of $G(\bi{m})$:
    $\langle\bi{x}\rangle = \arg\min_{\bi{m}}G(\bi{m})$. 
    Once the above Gibbs free energy is exactly calculated, the extremal conditions of $\bi{h}$ and $\bi{m}$ generally associate the average $\langle\bi{x}\rangle$ and local field \cite{P1982}.     
    However, the evaluation of equation (\ref{eq:Gibbs_raw}) is computationally difficult in general.
    To overcome this difficulty, we take the following expectation consistent or the adaptive TAP approach \cite{OW,OW2,OW3}.
    First, we define an alternative Gibbs free energy:
    \begin{equation}\label{eq:Gibbs_second_moment}
    \fl
    	G(\bi{m}, Q) \equiv \mathop{\rm extr}_{\bi{h}, \Lambda}\left[\bi{h}^\top \bi{m} - \frac{N}{2}\Lambda Q -\frac{1}{\beta}\ln\left\{\int e^{-\frac{\beta}{2}\|\bi{y} - A\bi{x}\|_2^2 +\beta \bi{h}^\top\bi{x} -\frac{\beta}{2}\Lambda \|\bi{x}\|_2^2 -\beta\lambda\|\bi{x}\|_1} d\bi{x} \right\}\right],
    \end{equation}
    which provides the constraints on the first and macroscopic second moments so that $\langle\bi{x}\rangle, \langle |\bi{x}|^2 \rangle/N = \arg\min_{\bi{m}, Q} G(\bi{m}, Q)$. 

	Unfortunately, equation (\ref{eq:Gibbs_second_moment}) is also difficult to evaluate directly.
    The adaptive TAP approach resorts this calculation to the following approximation:
	\begin{eqnarray}\label{eq:Gibbs_ada}
    \fl
    	G(\bi{m}, Q) \simeq \phi_{{\mathrm{ada}}}(\bi{m}, Q) \equiv \widetilde{\phi} (\bi{m}, Q; l = 0) 
        + \phi^{\mathrm{G}}(\bi{m}, Q; l=1) 
        - \phi^{\mathrm{G}}(\bi{m}, Q; l=0), \\
    \fl
            		\widetilde{\phi} (\bi{m}, Q; l) \equiv \mathop{\rm extr}_{\bi{h}, \Lambda} \left\{
            	\bi{h}^{\top}\bi{m} - \frac{N}{2}\Lambda Q 
                - \frac{1}{\beta}\ln \int e^{-\frac{\beta l}{2}\|\bi{y} - A\bi{x}\|_2^2 + \beta \bi{h}^\top \bi{x} - \frac{\beta}{2}\Lambda \|\bi{x}\|_2^2 -\beta\lambda \|\bi{x}\|_1}  d\bi{x}
            \right\}, \\ 
            \fl
            \phi^{\mathrm{G}}(\bi{m}, Q; l) \equiv \mathop{\rm extr}_{\bi{h}^{\mathrm{G}}, \Lambda^{\mathrm G}}\left\{
            \bi{h}_{\mathrm{G}}^{\top} \bi{m} - \frac{N}{2}\Lambda_{{\mathrm G }}Q 
            -\frac{1}{\beta}\ln \int e^{-\frac{\beta l }{2}\|\bi{y} - A \bi{x}\|_2^2+ \beta \bi{h}_G^\top \bi{x} -\frac{\beta}{2}\Lambda_G\|\bi{x}\|_2^2}  d\bi{x}
            \right\},
    \end{eqnarray}
	where $\widetilde{\phi}(\bi{m}, Q;l=0)$, $\phi^{\rm G}(\bi{m}, Q;l=1)$, and $\phi^{\rm G}(\bi{m}, Q;l=0)$ are the free energies for the modified distributions: the first term is a factorized distribution but contains the original non-Gaussian prior factor $e^{-\beta\lambda\|\bi{x}\|_1}$, while the second and third terms are the global and factorized multivariate Gaussian distribution that replaces the prior factor $e^{-\beta\lambda\|\bi{x}\|_1}$ with a Gaussian factor $e^{-\beta\Lambda_{\rm G}\|\bi{x}\|_2^2/2}$.
    In contrast to the original form of Gibbs free energy (\ref{eq:Gibbs_second_moment}), adaptive TAP free energy (\ref{eq:Gibbs_ada}) can be easily calculated as it is composed of only integration over the multivariate Gaussian and factorized distributions.
    The evaluation of the integrals and extremal conditions in the second and third terms of equation (\ref{eq:Gibbs_ada}) provides the following expression of $\phi_{\rm ada}$:
       \begin{eqnarray}
       \fl
          \phi_{\rm ada}(\bi{m}, Q) = \mathop{\rm extr}_{\bi{h}, \Lambda} \left[
              \frac{1}{2}\|\bi{y} - A\bi{m}\|_2^2- \frac{N\Lambda Q}{2}
              - \frac{N}{\beta} G(-\chi;J) \right. \nonumber\\
              \left.
              + \bi{h}^\top \bi{m} - \frac{1}{2\Lambda} \sum_{i=1}^N (|h_i|- \lambda)^2 \Theta(|h_i|-\lambda)
           \right],
       \end{eqnarray}
    where $\chi\equiv\beta(Q-q), q\equiv\sum_i m_i^2/N$.
    It has been shown \cite{OW2,KV2014} that the above free energy $\phi_{\mathrm{ada}}(\bi{m}, Q)$ is asymptotically consistent with replica theory in the sense that $\lim_{\beta\to\infty, N\to\infty}\Exp{\mathop{\rm extr}_{\bi{m}, Q}\phi_{\rm ada}(\bi{m}, Q)}_{A,\xi}/N = \Exp{f}_{A,\bi{\xi}}$ when $A$ is a sample from the rotationally invariant ensemble.
	Thus, the extremal condition on $\bi{h}, \Lambda, \bi{m}, Q$ and linear response argument give the intended TAP/cavity equations, which connect the local field and LASSO estimator for the current matrix ensembles for $\beta\to\infty,N\to\infty$:
    \begin{eqnarray}
        	\bi{h} = \Lambda \bi{m} + A^\top(\bi{y} - A\bi{m}), \label{eq:local_fields}\\
            m_i = \frac{h_i - \lambda \mathop{\rm sgn}(h_i)}{\Lambda}\Theta(|h_i| - \lambda), \\
            \Lambda = 2G'(-\chi), \label{eq:lambda}\\
            \chi = \frac{1}{N\Lambda} \sum_{i=1}^N\Theta(|h_i| - \lambda) = \frac{\varrho_{\rm active}}{\Lambda}, \label{eq:chi}
    \end{eqnarray}
    where $\varrho_{\rm active} \equiv \sum_{i=1}^N \Theta(|h_i| -\lambda)/N=|\{i|\widehat{x}^{\rm LASSO}_i\neq 0\}|/N$ is the active component density of the LASSO solution (\ref{eq:LASSO}).
    
    \subsubsection{General construction procedure of the de-biased estimator, confidence interval, and $p$-value:} \label{sec:method}
    In summary, once the LASSO estimator $\widehat{\bi{x}}^{\rm LASSO}(\bi{y},A;\lambda)$ is obtained for a set of $(\bi{y}, A; \lambda)$ by using versatile algorithms for the optimization problem (\ref{eq:LASSO}) such as least-angle regression \cite{EHJT2004}, coordinate descent \cite{FHT2010}, and approximate message passing \cite{DMM2009, R2011}, one can estimate the local fields $\bi{h}(\bi{y}, A; \lambda)$, de-biased estimator $\widehat{\bi{x}}^{\rm debiased}(\bi{y}, A; \lambda)$, confidence interval $\{\mathcal{I}_i(\alpha_{\rm sig})\}_i$, and $p$-value ${\rm P}_i$ as follows.
    We emphasize here that there is no need to use the derived TAP equation to obtain a LASSO estimator.
    
    First, the active component density $\varrho_{\rm active}$ is calculated from the LASSO solution:
    \begin{equation}
    	\varrho_{\rm active}(\bi{y}, A;\lambda) = \frac{1}{N}\left| \{i | \widehat{x}^{\rm LASSO}_i(\bi{y},A;\lambda)\neq 0\}\right|.
    \end{equation}
    Second, $z(-\chi)$ is obtained by combining equations (\ref{eq:chi_stieltjes}), (\ref{eq:G_prime}), (\ref{eq:lambda}), and (\ref{eq:chi}):
    $z(-\chi)$ is obtained as the solution of
    \begin{equation}
    	z = \frac{1-\varrho_{\rm active}}{\mathcal{S}_J(z)}.
    \end{equation}
	This equation is solved analytically or numerically depending on the cases. 
    Note that this equation is easily solved by using a simple iteration algorithm even if an analytical expression is not obtained.
    Then, $z^{\prime}(-\chi), \chi , G^{\prime}(-\chi;J), G^{\prime\prime}(-\chi;J)$, the Onsager coefficient $\widehat{Q}=\Lambda$, the local field $h(\bi{y},A;\lambda)$, the de-biased estimator $\widehat{\bi{x}}^{\rm debiased}(\bi{y},A;\lambda)$, the residual sum of squares, and the variance of the local field $\widehat{\chi}$ are obtained by subsequently substituting the obtained values into equations (\ref{eq:z_prime}), (\ref{eq:chi_stieltjes}), (\ref{eq:G_prime}), (\ref{eq:G_two_prime}), (\ref{eq:lambda}), (\ref{eq:local_fields}), (\ref{eq:debiased_estimator}), (\ref{eq:rss_def}), and (\ref{eq:chi_hat}):
    \begin{eqnarray}
        z^{\prime}(-\chi) = -\left[\int \frac{\rho_J(\lambda)}{(z(-\chi)-\lambda)^2}d\lambda\right]^{-1} \label{eq:z_prime_method},\\
        \chi = -\mathcal{S}_J(z(-\chi)), \label{eq:chi_stieltjes_method} \\
    	   	G^{\prime}(-\chi;J) = \frac{1}{2}\left(z(-\chi) + \frac{1}{\chi}\right), \label{eq:G_prime_method}\\
    G^{\prime\prime}(-\chi;J) = \frac{1}{2}\left(z^{\prime}(-\chi) + \frac{1}{\chi^2}\right),\label{eq:G_two_prime_method} \\
     \widehat{Q} = \Lambda = z(-\chi) + \frac{1}{\chi}, \label{eq:onsager_method}\\
		\bi{h}(\bi{y},A;\lambda) = \widehat{Q} \widehat{\bi{x}}^{\rm LASSO}(\bi{y},A;\lambda) + A^\top\left(\bi{y}-A\widehat{\bi{x}}^{\rm LASSO}(\bi{y},A;\lambda)\right)\label{eq:local_field_method}, \\
        \widehat{\bi{x}}^{\rm debiased}(\bi{y},A;\lambda) = \frac{\bi{h}(\bi{y},A;\lambda)}{\widehat{Q}},\label{eq:debiased_estimator_method}\\
        {\rm RSS} = \frac{1}{M}\left\|\bi{y} - A\widehat{\bi{x}}^{\rm LASSO}(\bi{y},A;\lambda) \right\|_2^2,\\
        \fl\widehat{\chi} = \frac{\gamma G''(-\chi; J)}{G'(-\chi; J) - G''(-\chi; J)\chi}\mathrm{RSS}  + \frac{-G''(-\chi; J)\gamma + 2\left(G'(-\chi; J)\right)^2}{G'(-\chi; J) - G''(-\chi; J)\chi}\sigma^2. \label{eq:chi_hat_method}
    \end{eqnarray}
  	Finally, the de-biased estimator's confidence interval $\{\mathcal{I}_i(\alpha_{\rm sig})\}_i$ and $p$-value $\{{\rm P}_i\}_i$ are obtained based on equations (\ref{eq:p-value})--(\ref{eq:confidence_interval}).

	Note that a consistent estimator of the error variance $\sigma^2$ should be needed when it is unknown.

\section{Numerical Experiment} \label{sec:num4} 
     
    \subsection{Settings}
    We perform numerical experiments to assess the usefulness of the proposed scheme.
    For this, we artificially generate the true sparse parameter $\bi{x}_0$, observation matrix $A$, and measurement noise $\bi{\xi}$.
    The true sparse parameter $\bi{x}_0$ is generated from the Bernoulli--Gauss distribution: $x_{0,i}\sim_{\rm i.i.d.} (1-\varrho)\delta(x_{0,i}) + \varrho \mathcal{N}(0,1)$ for $i=1,2,...,N$.
    The measurement noise is distributed according to the Gaussian distribution $\bi{\xi}\sim\mathcal{N}(0_M, \sigma^2 I_M)$.
    For the random observation matrix ensembles, the following ensembles are considered.
    \begin{enumerate}
	    \item The random i.i.d. Gaussian ensemble in which all entries of $A$ are i.i.d. Gaussian variables with mean $0$ and variance $1/N$. For this ensemble, the asymptotic eigenvalue distribution is given as the Marchenko-Pastur distribution \cite{MP1967}: 
        \begin{eqnarray}
        \rho(s) = (1-\gamma)\delta(s) + \frac{\gamma}{2\pi}\frac{\sqrt{(\lambda_+ - s)(s-\lambda_-)}}{s}\mathbb{I}_{[\lambda_-, \lambda_+]}(s), \\
        \lambda_\pm = \left(1\pm \sqrt{\gamma}\right)^2 , \\
       \mathbb{I}_S(x) = \cases{1&if $x\in S$\\
      0&otherwise \\}.
        \end{eqnarray}
         Then, the form of $G^\prime(-\chi;J), G^{\prime\prime}(-\chi;J), \chi$ and $\widehat{Q}$ are given as follows:
        \begin{eqnarray}
            G^\prime(-\chi;J) = \frac{\gamma}{2}\frac{1}{1+\chi}, \\
            G^{\prime\prime}(-\chi;J) = \frac{\gamma}{2}\frac{1}{(1+\chi)^2}, \\
            \chi = \frac{\varrho_{\rm active}}{\gamma - \varrho_{\rm active}},\\
            \widehat{Q} = \gamma - \varrho_{\rm active}. \label{eq:Lambda_Gauss}
        \end{eqnarray}
        By substituting the above expressions of $G^\prime, G^{\prime\prime}$ into (\ref{eq:chi_hat}), one can show that $\widehat{\chi}$ does not depend on $\sigma^2$. 
        This is the characteristic property of this ensemble.
        Generally, $\widehat{\chi}$ depends on the measurement noise $\sigma^2$.
        
        \item The row-orthogonal ensemble \cite{KV2014,VKC2016} constructed by randomly selecting $M$ rows from a randomly generated $N\times N$ orthogonal matrix. For this ensemble, the asymptotic eigenvalue distribution is given as $\rho(s) = (1-\gamma)\delta(s) + \gamma\delta(s-1)$. In this case, the form of $G^\prime(-\chi;J), G^{\prime\prime}(-\chi;J), \chi$ and $\widehat{Q}$ are given as follows:
        \begin{eqnarray}
            G^\prime(-\chi;J) = \frac{1}{2}\left(z(-\chi) + \frac{1}{\chi} \right),\\
            G^{\prime\prime}(-\chi;J) = \frac{1}{2}\left(z^\prime(-\chi) + \frac{1}{\chi^2} \right),\\
            \chi = \frac{\rho_{\rm A}(1-\varrho_{\rm active})}{\gamma - \varrho_{\rm active}},\\
            \widehat{Q} = \frac{\gamma - \varrho_{\rm active}}{1-\varrho_{\rm active}}, \label{eq:row_orthogonal}
          \end{eqnarray}
          where 
          \begin{eqnarray}
            z(-\chi) = - \frac{1- \chi + \sqrt{(\chi+1)^2- 4\gamma \chi} }{2 \chi},\\
            z^\prime(-\chi) = - \frac{1- 2 \gamma \chi + \chi + \sqrt{(\chi+1)^2- 4 \gamma \chi}}{2 \chi^{2} \sqrt{(\chi + 1)^2 - 4 \gamma \chi}}.
        \end{eqnarray} 

        \item The random discrete cosine transform (DCT) ensemble in which $A$ is constructed by randomly selecting $M$ rows from $N\times N$ DCT matrix.
       While this ensemble shares the same eigenvalue distribution as the row-orthogonal one, it is much more relevant for practical purposes, as the computational cost for observation and inference can be significantly reduced by using the fast Fourier transform technique. 
        In addition, although the rotationally invariant assumption on $O$ does not hold, this ensemble is also compatible with the current adaptive TAP scheme, as pointed by \cite{CO2018}.
        
        \item The geometric setup \cite{VKC2016, RFSK2017} in which $A$ is constructed as $A=U\Sigma V^\top$, where $U\in\mathbb{R}^{M\times M}$ and $V\in\mathbb{R}^{N\times N}$ are random samples from the uniform distribution of orthogonal matrices, and $\Sigma\in \mathbb{R}^{M\times N}$ is a diagonal matrix whose $(i,i)$-th element is given by $\nu_i\propto \tau^{i-1}$ for $i=1,2,...,M$. 
        The parameter $\tau \in (0,1]$ is chosen so that the given value of the peak-to-average eigenvalue ratio 
        \begin{equation}
        	\kappa \equiv \frac{\nu_1^2}{M^{-1}\sum_{i=1}^M\nu_i^2}
        \end{equation}
        is met and the singular values are scaled to satisfy the power constraint $1 = \frac{1}{N}\sum_{i=1}^M \nu_i^2$.
        The asymptotic eigenvalue distribution is given as 
        \begin{equation}
        	\rho(s)=(1-\gamma)\delta(s) + \frac{\gamma}{\eta s}\mathbb{I}_{(Be^{-\eta}, B]}(s),
        \end{equation}
        where $\eta$ and $B$ are related to the peak-to-average ratio $\kappa$:
        \begin{eqnarray}
        	\kappa = \frac{\eta}{1-e^{-\eta}}, \\
            B = \frac{\kappa}{\gamma}.
        \end{eqnarray}
        
        In this case, the explicit form of $G^\prime ,G^{\prime\prime}$ cannot be obtained.
        Thus, it should be evaluated numerically.
        To achieve this aim, we conduct the procedure explained in section \ref{sec:method}, using the expression of the Stieltjes transform and $z^{\prime}(-\chi)$:
        \begin{eqnarray}
          \fl
          \chi = -\mathcal{S}_J(z(-\chi))=-\int\frac{\rho(s)}{z(-\chi)-s}d\lambda = -\frac{1}{z(-\chi)}\left[1-\frac{\alpha}{\eta}\ln\frac{z(-\chi)-B}{z(-\chi)-Be^{-\eta}}\right], \\
          \fl 
          z^{\prime}(-\chi) = \frac{z(-\chi)^2}{-1+\frac{\gamma}{\eta}\ln \frac{z(-\chi)-B}{z(-\chi)-Be^{-\eta}} -\frac{z(-\chi)}{(z(-\chi)-Be^{-\eta})(z(-\chi)-B)}}.
        \end{eqnarray}
        
    \end{enumerate}
    
    We mainly use the random i.i.d. Gaussian ensemble and random DCT ensemble for the numerical experiments. The geometric setup is only used in section \ref{subsubsection:hyper_parameter_selection}. We do not use the original row-orthogonal setup.
    
	Once a tuple of $(\bi{x}_0, A, \bi{\xi})$ is generated, we calculate $\widehat{\bi{x}}^{\rm LASSO}$, $\bi{h}$, $\chi, \widehat{\chi}$ and $\widehat{Q} = \Lambda$, $\widehat{\bi{x}}^{\rm debiased}$ by using the procedure explained in section \ref{sec:method}.
 	   \begin{figure}[b]
            \centerline{\includegraphics[width=16cm]{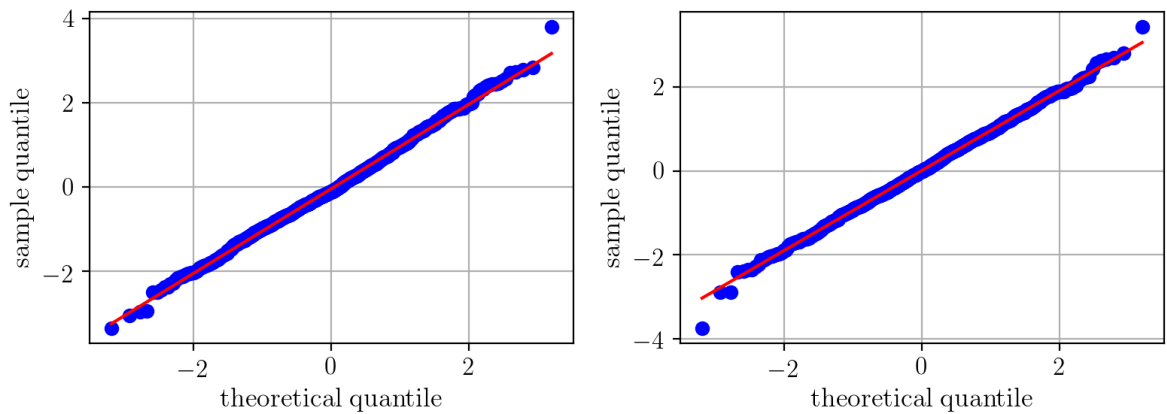}} 
           \caption{$Q$-$Q$ plot of $\{(h_i-x_{0,i}\widehat{Q})/\sqrt{\widehat{\chi}}\}_i$. 
           The red line is the unit slope and zero intercept line.
           Left: The i.i.d. Gaussian case. Right: The random DCT case. 
           }
           \label{fig:fig2}
        \end{figure}
	To estimate the error variance $\sigma^2$ needed in the random DCT case, we use the naive cross-validation-based estimator:
    \begin{equation}
    	\hat{\sigma}^2(\bi{y}, A;\hat{\lambda}) \equiv \frac{1}{M - N \varrho_{\rm active}} \left\|\bi{y} - A\widehat{\bi{x}}^{\rm LASSO}(\bi{y}, A;\hat{\lambda})\right\|_2^2,
    \end{equation}
    where $\hat{\lambda}$ is selected by $K$-fold cross-validation.
	In \cite{RTF2013}, it is empirically shown that this estimator robustly estimates the error variance, more so than its competitors.
    
    We use $N_{\rm s} = 1000$ different sets of pairs $(A, \bi{\xi})$ for fixed $\bi{x}_0$ to evaluate the statistical properties of the observables.
    We set $\varrho=0.1, \gamma=0.5, \sigma^2=0.02$, and $K=40$, except for the geometric setup. 
    In the geometric setup, we set $\varrho =0.1, \gamma=0.8, \sigma^2=0.02$, and $\kappa=8$.

    \subsection{Results}
    	\subsubsection{Distribution of the local fields and de-biased estimators:}
\begin{figure}[t]
        \centerline{\includegraphics[width=16cm]{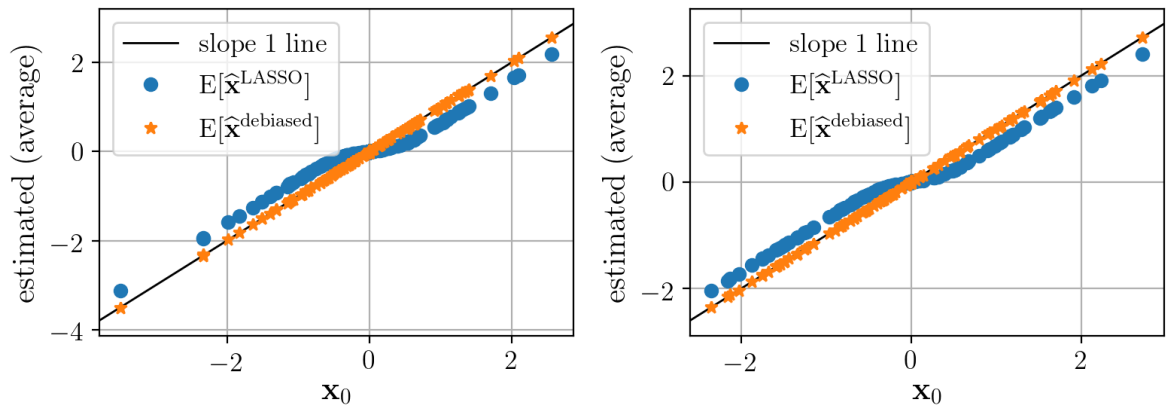}} 
       \caption{De-biasing effect of $\widehat{\bi{x}}^{\rm debiased}$.　The blue points stand for the average value of the LASSO solution $\widehat{\bi{x}}^{\rm LASSO}$ and orange points stand for the average value of the de-biased estimator $\widehat{\bi{x}}^{\rm debiased}$. The black line is the unit slope and zero intercept line.
       Left: The i.i.d. Gaussian ensemble case. Right: The random DCT ensemble case.}
       \label{fig:fig3}
    \end{figure}

\begin{figure}[t]
        \centerline{\includegraphics[width=16cm]{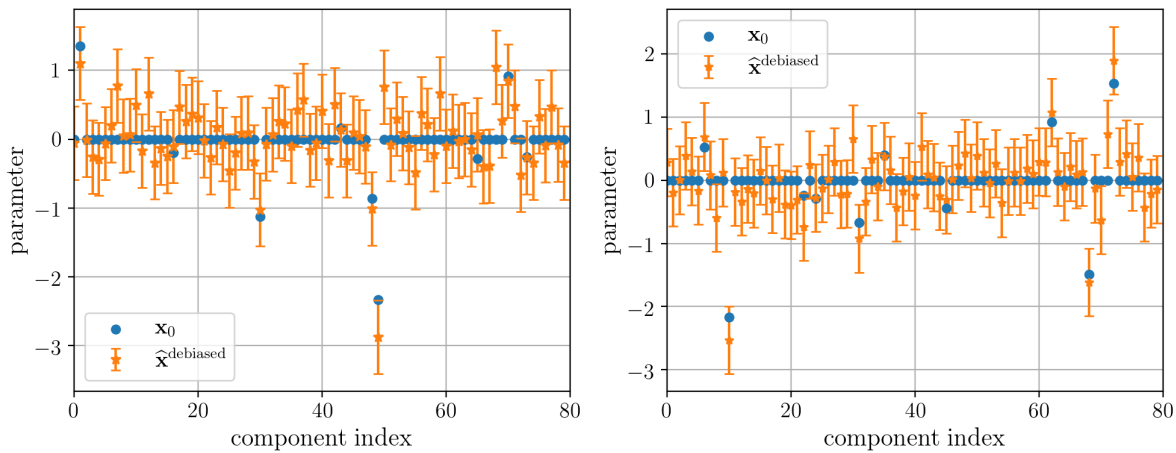}} 
       \caption{Constructed de-biased estimator $\widehat{\bi{x}}^{\rm debiased}$ and its $95\%$ confidence interval. In both the left and the right panels, the blue points stand for the true parameter $\bi{x}_0$ and orange points are the de-biased estimator $\widehat{\bi{x}}^{\rm debiased}$. The orange error bars are the $95\%$ confidence intervals.
       Left: The i.i.d. Gaussian ensemble case. Right: The random DCT ensemble case.}
       \label{fig:fig4}
    \end{figure}

\begin{figure}[t]
        \centerline{\includegraphics[width=16cm]{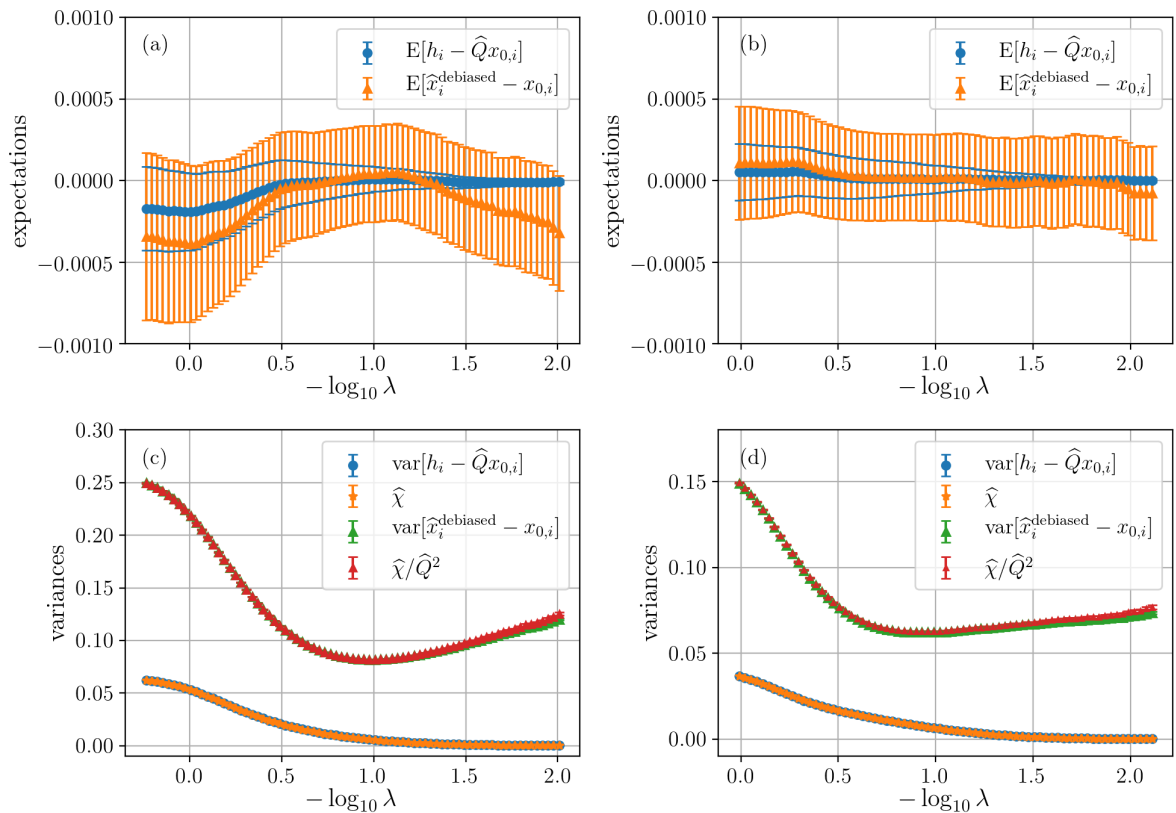}}  
       \caption{
       (a) and (b): Mean of $\{h_i - \widehat{Q}x_{0,i}\}_i$ and $\{\widehat{x}^{\rm debiased} - x_{0,i}\}_i$.
       (c) and (d): Comparison of the estimated and empirical values of the variances of $\{ h_i-\widehat{Q} x_{0,i}\}_i$ and $\{\widehat{x}^{\rm debiased}_i  -x_{0,i}\}_i$.
      The orange and red points represent the theoretically estimated values. The blue and green points stand for the empirical ones.
       
       (a) and (c) are the i.i.d. Gaussian case. (b) and (d) are the random DCT case.
       }
       \label{fig:fig5}
    \end{figure}

        First, we examine the statistical properties of the local fields and de-biased estimators.
        Figure \ref{fig:fig2} plots the sample quantiles of $\{(h_i - \widehat{Q}x_{0,i})/\sqrt{\widehat{\chi}}\}_i$ versus the theoretical quantiles of the standard normal distribution for one configuration of $(\bi{x}_0, A,\bi{\xi})$.
        It is clear that all the points are close to the line with unit slope and zero intercept.
        Further, Figure \ref{fig:fig3} plots the average values of the $\widehat{\bi{x}}^{\rm LASSO}$ and $\widehat{\bi{x}}^{\rm debiased}$ versus the true parameter $\bi{x}_0$.
        In contrast to the LASSO estimators, which are shrunk toward zero by the regularization term, $\widehat{\bi{x}}^{\rm debiased}$ efficiently reduces the LASSO estimator's bias. 
        The average is taken over $N_s$ realizations of $(A,\bi{\xi})$.
        These results validate our theoretical predictions on the local fields and de-biased estimators.
        Figure \ref{fig:fig4} plots the constructed de-biased estimators and their $95\%$ confidence intervals.
        We show only the first $80$ components for the sake of clarity.
     Although Figures \ref{fig:fig2}--\ref{fig:fig4} show the results for one value of $\lambda$, 
     the same results are obtained for a wide range of $\lambda$.
     The means of $\{h_i - \widehat{Q}x_{0,i}\}_i$ and $\{\widehat{x}^{\rm debiased} - x_{0,i}\}_i$ are zero in both the i.i.d. Gaussian and the random DCT cases (Figure \ref{fig:fig5} (a) and (b)).
     Further, the variances of $\{ h_i-\widehat{Q} x_{0,i}\}_i$ and $\{\widehat{x}^{\rm debiased}_i -x_{0,i}\}_i$ agree with their estimates of $\widehat{\chi}$ and $\widehat{\chi}/\widehat{Q}^2$, respectively for the whole range of the weight of the $L_1$ regularizer $\lambda$ (Figure \ref{fig:fig5} (c) and (d)).
     
     \subsubsection{Hypothesis testing:}
    An important advantage of the proposed scheme over LASSO is that it provides a hypothesis testing method with a null hypothesis that a certain parameter vanishes.
    Although LASSO provides a parameter selection rule that selects an active component set $\mathcal{A}(\bi{y}, A;\lambda)$ as $\mathcal{A}(\bi{y}, A;\lambda)=\{i | \widehat{x}^{\rm LASSO}_i(\bi{y}, A;\lambda)\neq0 \}$, it cannot measure the statistical significance for finding an active component.
    
    \begin{figure}[t]
        \centerline{\includegraphics[width=16cm]{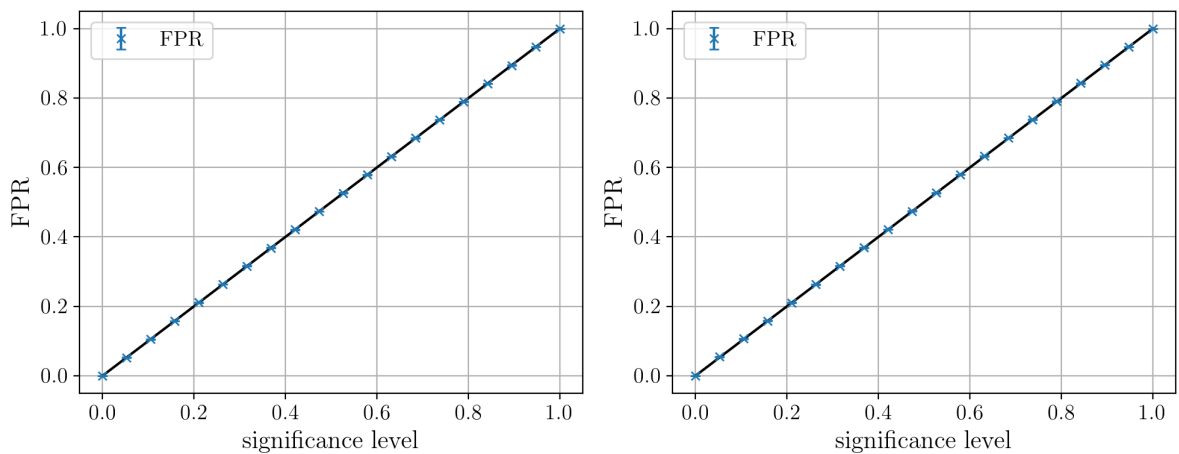}} 
       \caption{Significance level versus the observed false positive rate (FPR). The black solid line is the unit slope line. Left: the i.i.d. Gaussian case. Right: the random DCT case.}
       \label{fig:fig6}
    \end{figure}

    \begin{figure}[b]
        \centerline{\includegraphics[width=16cm]{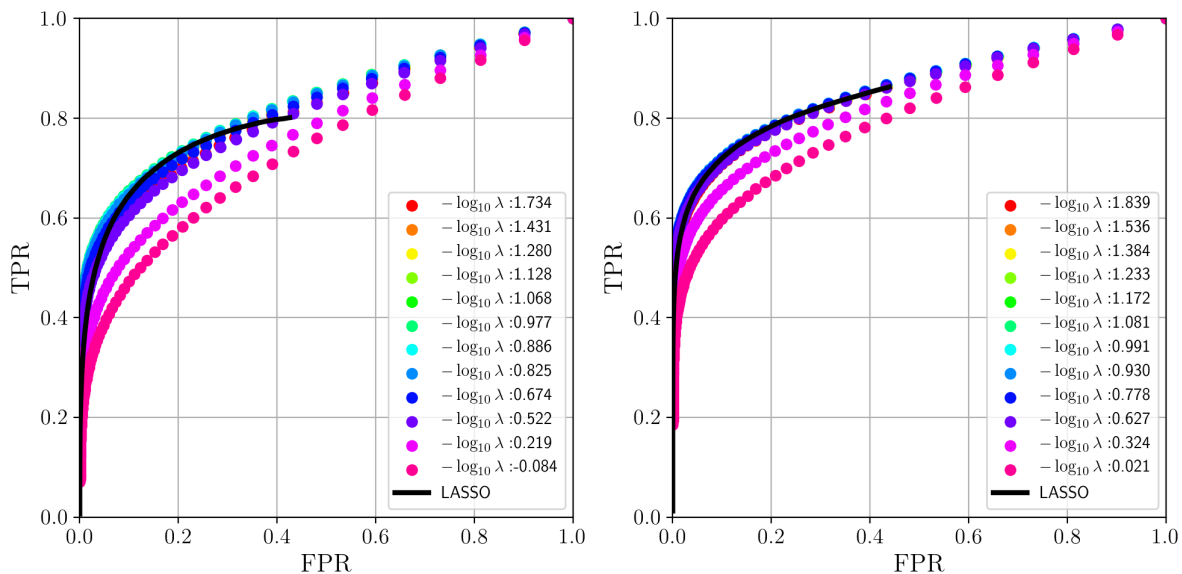}} 
       \caption{ROC (receiver operating characteristic) curves. The black solid line represents the ROC curve for LASSO obtained by varying the regularization strength $\lambda$. The points correspond to the ROC curve for the proposed hypothesis testing method. Left: the i.i.d. Gaussian case. Right: the random DCT case.}
       \label{fig:fig7}
    \end{figure}
    
    Specifically, we are interested in testing an individual null hypothesis $H_{0,i}: x_{0,i}=0$ versus the alternative hypothesis $H_{1,i} : x_{0,i}\neq 0$, assigning a $p$-value of ${\rm P}_i$ for these tests.
    To this end, we evaluate the $p$-value of $\{{\rm P}_i\}$ by using equation (\ref{eq:p-value}) for a two-tailed test.
    Then, the decision rule is to reject the null hypothesis $H_{0,i}$ if the observed $p$-value ${\rm P}_i$ is lower than $\tilde{\alpha}_{\rm sig}$ and to accept the alternative hypothesis otherwise:
    \begin{equation}
      \label{eq:hypothesis_testing_procedure}
      \widehat{T}_i(\bi{y}, A;\lambda) = \cases{1&if ${\rm P}_i \leq \tilde{\alpha}_{\rm sig} $ (reject)\\
      0&otherwise (accept)\\},
	\end{equation}
    where $\tilde{\alpha}_{\rm sig}$ is the significance level.
    We use $\widehat{T}$ as a rejection flag.
    This procedure ensures that the type I error probability or the FPR is $\tilde{\alpha}_{\rm sig}$.
    Here, the FPR is the probability of falsely rejecting the null hypothesis $H_{0,i}$:
    \begin{equation}
    	\mathrm{FPR}\equiv
        \frac{\left| \left\{i | \widehat{T}_i =1\;\mathrm{and}\;x_{0,i}=0\right\}\right| }{\left| \left\{i | x_{0,i}=0\right\}\right|}.
    \end{equation}
    Indeed, Figure \ref{fig:fig6} shows that the significance level $\tilde{\alpha}_{\rm sig}$ and empirical true positive rate (TPR) are in excellent agreement.
    
    Further, we examine the relation between the FPR and TPR or the statistical power achieved by LASSO and our hypothesis testing procedure.
    Here, the TPR is the probability that the test correctly rejects the null hypothesis $H_{0,i}$:
    \begin{equation}
    	{\rm TPR} \equiv 
        \frac{\left| \left\{i | \widehat{T}_i =1\;\mathrm{and}\;x_{0,i}\neq 0\right\}\right| }{\left| \left\{i | x_{0,i}\neq 0\right\}\right|}.
    \end{equation}
    Note that although we can control the FPR by varying the significance level $\tilde{\alpha}_{\rm sig}$, the TPR cannot be controlled. 
    Thus, a performance measure of the variable selection procedure by hypothesis testing can be given as the TPR for each value of the FPR.
    We evaluate the performance of hypothesis testing by using the ROC curve, which plots the TPR versus the FPR as an implicit function of $\tilde{\alpha}_{\rm sig}$.
    We examine the TPR and FPR by varying the significance level $\tilde{\alpha}_{\rm sig}$ for each regularization parameter $\lambda$.
    For comparison purposes, we also plot the ROC curve for LASSO.
    For LASSO, the TPR and FPR are examined by changing the regularization parameter $\lambda$.
    
    \begin{figure}[b]
        \centerline{\includegraphics[width=16cm]{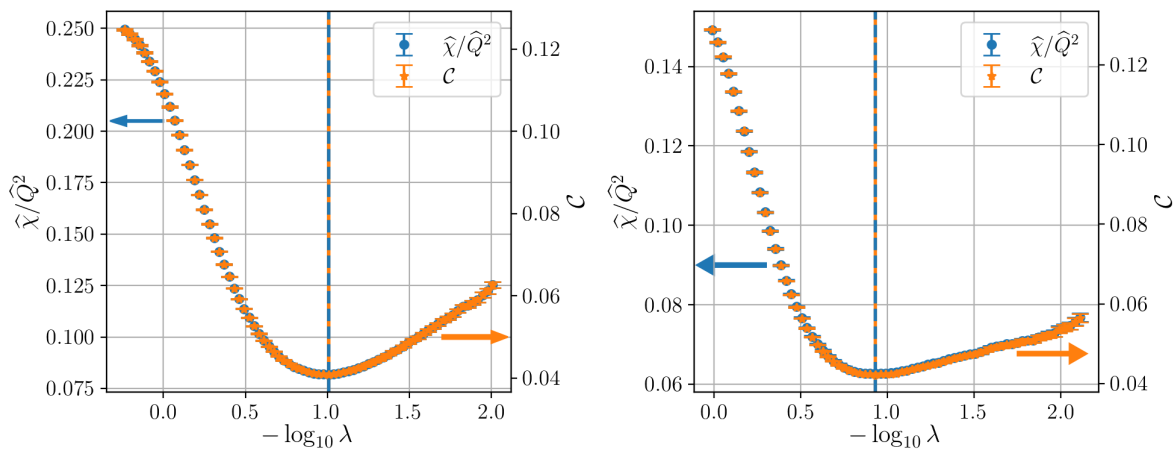}} 
       \caption{Comparison of the width of the confidence interval versus the leave-one-out cross-validation error. The blue and orange points show the average width of the confidence interval and leave-one-out cross-validation error, respectively. 
       The blue solid line and orange dashed line indicate the value of $\lambda$ that minimizes the confidence interval and leave-one-out cross-validation error, respectively.
       The left and right vertical axes represent the values of $\widehat{\chi}/\widehat{Q}^2$ and $\mathcal{C}$, respectively.
       The axis range for $\mathcal{C}$ is chosen according to equation (\ref{eq:variance_linear}) so that the curves of $\widehat{\chi}/\widehat{Q}^2$ and $\mathcal{C}$ overlap.
       The values of $\lambda$ that minimize the width of the confidence interval and cross-validation error perfectly coincide as expected.
       Left: The i.i.d. Gaussian case. Right: The random DCT case.}
       \label{fig:fig8}
    \end{figure}

    Figure \ref{fig:fig7} summarizes the results averaged over $N_s$ configurations of $(A,\bi{\xi})$.
    It is observed that for some values of $\lambda$ around which the variance of the de-biased estimator is minimized, our testing procedure performs slightly better than LASSO in the sense that the TPR of the testing method is slightly larger than that of LASSO's one for some values of the FPR.
    In the case of LASSO, when the measurement ratio $\gamma$ is sufficiently small, the TPR and FPR do not coincide with $(1, 1)$ for finite $\lambda>0$, as the consistency property does not hold in such a situation and the number of active components of the LASSO estimator is always smaller than $\min(N,M)$ \cite{BS2011}.
    On the contrary, as our hypothesis testing procedure always approaches the point $(1,1)$ from $(0,0)$, we can examine the TPR for all the values of the FPR $\in[0,1]$.
    The superiority of the TPR comes from the fact that we are using the knowledge of the ensemble of the observation matrix.
    Further, as the hypothesis testing procedure controls the FPR and TPR by varying the significance $\alpha_{\rm sig}$ but not $\lambda$, one does not suffer from the shrinkage effect in the variable selection procedure.
    This is another advantage over variable selection by LASSO.
    These observations show the utility of our hypothesis testing procedure.

    \begin{figure}[b]
        \centerline{\includegraphics[width=16cm]{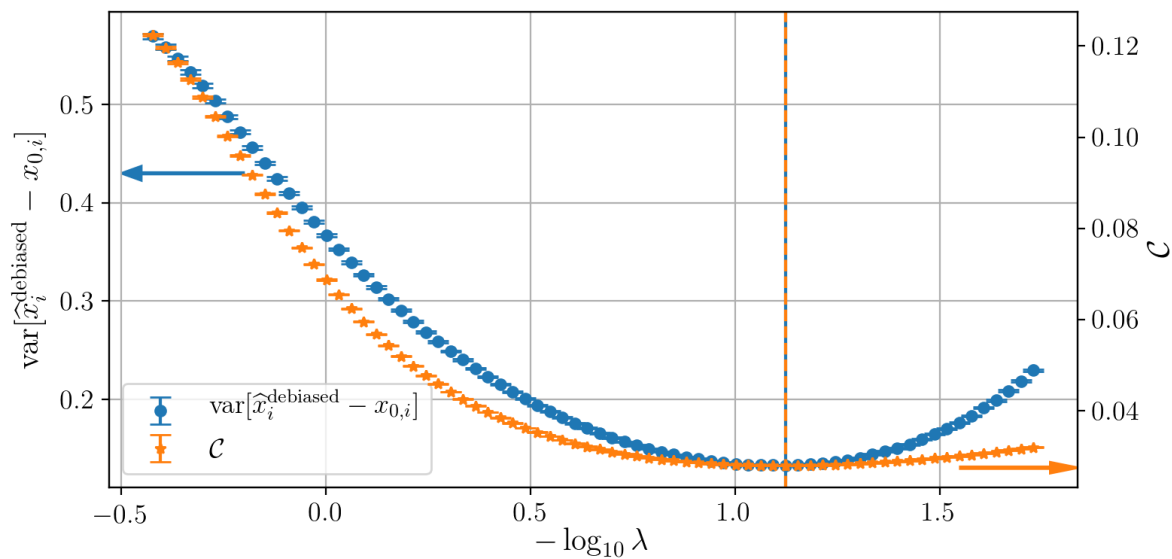}} 
       \caption{Comparison of the width of the confidence interval versus the leave-one-out cross-validation error for the geometric case.
       Here, $\widehat{\chi}/\widehat{Q}^2$ is evaluated by var$\left[\widehat{x}^{\mathrm{debiased}}_i-x_{0,i}\right]$.
       The blue and orange points show the average width of the confidence interval and leave-one-out cross-validation error, respectively. 
       The blue solid line and orange dashed line indicate the value of $\lambda$ that minimizes the confidence interval and leave-one-out cross-validation error, respectively.
       The left and right vertical axes represent the values of var$\left[\widehat{x}^{\mathrm{debiased}}_i-x_{0,i}\right]$ and $\mathcal{C}$, respectively.
       Unexpectedly, the values of $\lambda$ that minimize the width of the confidence interval and cross-validation error perfectly coincide.}
       \label{fig:fig9}
    \end{figure}

		\subsubsection{Hyperparameter selection via confidence interval minimization:}\label{subsubsection:hyper_parameter_selection}
        The issue of hyperparameter selection is noteworthy here.
        As LASSO has the hyperparameter $\lambda$ that controls the strength of the regularization, one should choose a value of $\lambda$ based on some criteria.
        As shown in Figure \ref{fig:fig5}, the estimated variance of the de-biased estimator $\widehat{\chi}/\widehat{Q}^2$ has a minimum value at some $\lambda>0$.
        At this point, one can estimate $\bi{x}_0$ with the highest conviction in the sense that the confidence interval has the smallest width.
        It is therefore expected that the estimated variance of the de-biased estimator itself serves as a hyperparameter selection criterion.
        Indeed, in the i.i.d. Gaussian and row-orthogonal/random DCT cases, one can analytically show that minimizing the confidence interval is the equivalent to the minimization of the leave-one-out cross-validation error $\mathcal{C}$:
        \begin{eqnarray}
            \label{eq:variance_linear}
            \fl
            \frac{\widehat{\chi}}{\widehat{Q}^2} = \cases{\frac{1}{\gamma}\mathcal{C}& the i.i.d. Gaussian case, \\
            \frac{1-\gamma}{\gamma}\mathcal{C} + \sigma^2 & the row-orthogonal or the random DCT cases. \\}
        \end{eqnarray}
        In other words, the leave-one-out cross-validation error and width of the confidence intervals are related with the linear transformation in these cases (Figure \ref{fig:fig8}).

        Here, the leave-one-out cross-validation error is a widely used hyperparameter selection criterion that evaluates prediction performance, defined as follows:
        \begin{equation}
            \mathcal{C}(\bi{y}, A;\lambda) = \frac{1}{M}\sum_{i = 1}^{M}\frac{1}{2}\left\| y_i -  \bi{a}_i^\top \widehat{\bi{x}}^{\rm LASSO}(\bi{y}_{\backslash i}, A_{\backslash i};\lambda)\right\|_2^2, 
        \end{equation}
        where the symbol $\backslash i$ denotes the absence of the $i$-th component (e.g., $\bi{a}_{\backslash i}=(a_1,...,a_{i-1},a_{i+1},...,a_N)^\top$) and each term in the summation evaluates the fitness to the $i$-th data when the true signal is inferred from the other data.
        In the settings considered here, the above leave-one-out cross-validation error is expressed as follows \cite{OK2016,RM2018}:
        \begin{equation}
            \mathcal{C} = \left(1-\frac{\varrho_{\rm active}}{\gamma}\right)^{-2} {\rm RSS} = \left(1-\frac{2\chi G^\prime(-\chi;J)}{\gamma}\right)^{-2}{\rm RSS}.
            \label{eq:looe}
        \end{equation}
        By substituting the expression of the leave-one-out cross-validation error (\ref{eq:looe}) into equation (\ref{eq:chi_hat}), the relations (\ref{eq:variance_linear}) are obtained.

        To investigate the validity of the above observation that the confidence interval minimization and leave-one-out cross-validation error minimization provide the same $\lambda$, we test the geometric setup case in which $\widehat{\chi}/\widehat{Q}^2$ is not expressed as a linear function of $\mathcal{C}$.
        Figure \ref{fig:fig9} compares the variance of $\{\widehat{x}^{\rm debiased} - x_{0,i}\}_i$ with the leave-one-out cross-validation error (\ref{eq:looe}).
        Surprisingly, the minimization of these two quantities seems to provide the same value of $\lambda$, although they do not have a functional relation as (\ref{eq:variance_linear}).

        From the above observations, we speculate that the minimization of the confidence interval proposed here and the minimization of the leave-one-out cross-validation error yields the same value of $\lambda$ for LASSO in general rotationally invariant observation matrices, but further investigation in this direction is still needed.

\section{Summary}
	We developed a new computationally feasible scheme for de-biasing and uncertainty estimation in LASSO in the case of rotationally invariant observation matrix ensembles and validated the proposed scheme by using numerical experiments.
    We focused on the development of a de-biased estimator that has a confidence interval and hypothesis testing scheme for the null hypothesis that a certain parameter vanishes.
	The numerical experiments showed that the proposed method efficiently constructed de-biased estimators with confidence intervals and $p$-values for the intended hypothesis testing. 
    We revealed that the proposed hypothesis testing slightly improved the variable selection performance in the sense that the TPR of the testing method achieves a slightly larger value than that of the LASSO's one for some values of the FPR.
    Further, we examined the utility of the estimator of the confidence interval as a criterion for determining the hyperparameter. Surprisingly, minimizing the width of the confidence interval was equivalent to the minimization of the leave-one-out cross-validation error in our investigation.

    Although we only focused on LASSO for linear models, future work could include an extension to other sparse regression methods such as the elastic net \cite{ZH2005} as well as generalized linear models.

\section*{Acknowledgement}
Support by JSPS KAKENHI Nos. 25120013 and 17H00764 (YK) is acknowledged. 

\appendix
\section{Derivation of the Free Energy Density} \label{app:num1} 
	To take the average that appears in (\ref{eq:free_energy_density}), we use the replica method \cite{MPV1987} based on the identity for $n\in\mathbb{R}$:
    \begin{equation}\label{eq:a1}
    	f = -\lim_{\beta\to\infty}\lim_{N\to\infty}\frac{1}{\beta N}\lim_{n\to 0}\frac{\Exp{Z^n}_{A,\bi{\xi}}}{n}.
    \end{equation}
    In the replica method, we first take the average of the $n$-th power of the partition function over the randomness of $A, \bi{\xi}$ for the positive integer $n\in\mathbb{N}$, and then analytically continue the obtained expression to real $n\in\mathbb{R}$ to take the limit $n\to0$, exchanging the order of the limits.
    
    For the general matrix ensembles considered here, it is convenient to first take the average over $\bi{\xi}$.
    By taking this average, we obtain the following expression under the replica symmetric ansatz:
    \begin{equation}\label{eq:a2}
          \Exp{Z^n}_{A,\bi{\xi}}  = \int \Exp{\exp\left( \frac{1}{2} \mathrm{Tr} J L \right)}_A  e^S dQdqdm,
      \end{equation}
      where $L$, $\bi{u}_{a}$, and $S$ are defined as follows:
      \begin{eqnarray}
          L \equiv \frac{\beta^2\sigma^2}{ 1 + \beta n\sigma^2}\left(\sum_a \bi{u}_a\right)\left(\sum_a \bi{u}_a\right)^\top
          -\beta \sum_a \bi{u}_a \bi{u}_a^\top, \\
          \bi{u}_a \equiv \bi{x}_a - \bi{x}_0, \\
    \fl e^S \equiv \int \prod_{a=1}^n\delta(NQ - \bi{x}_a^\top \bi{x}_a)
      \delta(Nm - \bi{x}_a^\top \bi{x}_0) \nonumber \\
      \times\prod_{1\leq a<b\leq n}\delta(Nq - \bi{x}_a^\top \bi{x}_b)\exp\left\{-\frac{N\gamma}{2}\beta n\sigma^2 -\beta\lambda\sum_a\|\bi{x}_a\|_1 \right\}d\bi{x},
    \end{eqnarray}
    where $\bi{u}_a$ and $\bi{x}_a$ are the $a$-th replica's variable.
    In \cite{MPR1994}, it was shown that under the rotational invariance assumption on the random matrix $J=A^\top A$ for eigenvalue decomposition $J=ODO^\top$ considered in this study, the average over $A$ that appears in equation (\ref{eq:a2}) is evaluated by using the eigenvalues $\{s_i\}_i$ of $L/N$ for sufficiently large $N$:
    \begin{equation}
    	\Exp{\exp\left( \frac{1}{2} \mathrm{Tr} J L \right)}_A = \exp\left\{N \sum_{i}G(s_i; J) \right\},
    \end{equation}
    where $G(x;J)$ is the function defined in (\ref{eq:def_G}).
    Under the replica symmetric ansatz, $L/N$ has three types of eigenvalues: $s_1=\beta \Delta Q - \beta n (q- 2m + \varrho) + n \beta^2 \sigma^2 \Delta Q$, $s_2=-\beta \Delta Q$, and $s_3 =0$. 
    The number of degeneracy is $1, n-1$, and $N-n$, respectively.
    Thus, we obtain the following expression up to the leading order in $n$:
    \begin{eqnarray}\label{eq:a_energy}
    \fl
    	\Exp{e^{ \frac{1}{2} \mathrm{Tr} J L }}_A 
        = \exp\left[-Nn\beta\left\{-G(-\beta \Delta Q; J)/\beta \right.\right.\nonumber \\
     \left.\left.+ G'(-\beta \Delta Q; J)(q-2m + \varrho - \beta\Delta Q \sigma^2) \right\}\right].
    \end{eqnarray}
    
    On the contrary, by using the Fourier transform of the delta function and Hubbard--Stratonovich transform: $e^{B^2/2A}=\int e^{-Ax^2/2 + Bx}\sqrt{\frac{A}{2\pi}}dx$ for $A>0,B\in\mathbb{R}$, the factor $e^S$ is given as follows:
    \begin{eqnarray}
    \fl
    	   e^{S} = \int \exp\left[Nn\left\{ \frac{\gamma\sigma^2}{2} + 
           \frac{q\widetilde{q}}{2} + \frac{Q\widetilde{Q}}{2} - m\widetilde{m}\right.\right. \nonumber \\
           \left.\left. +\frac{1}{N}\sum_{i=1}^N \int \ln \phi(x_{0,i}, z_i, \widetilde{Q},\widetilde{q}, \widetilde{m};\beta, \lambda)Dz_i
           \right\}
        \right]d\widetilde{Q}d\widetilde{q}d\widetilde{m},\label{eq:a_entropy_1}\\
    \fl
         \phi(x_{0,i}, z_i, \widetilde{Q},\widetilde{q}, \widetilde{m};\beta, \lambda) =
         \int \exp\left\{-\frac{\widetilde{Q} + \widetilde{q}}{2}x_i^2 + (\widetilde{m}x_{0,i} + \sqrt{\widetilde{q}}z_i)x_i - \beta\lambda |x_i| \right\}dx_i.\label{eq:a_entropy_2}
    \end{eqnarray}
    For $\beta\to\infty$, the relevant variables scale as $\beta(Q-q)=\chi, \widetilde{Q} + \widetilde{q} = \beta\widehat{Q}, \widetilde{m}=\beta\widehat{m}$, and $\widetilde{q}=\beta^2\widehat{\chi}$ of order unity to ensure an appropriate limit $f$ exists.
    Finally, by combining equations (\ref{eq:a_energy}--\ref{eq:a_entropy_2}) and evaluating the integrals by adopting the saddle point method, we obtain equation (\ref{eq:free_energy_density_saddle}) for $\beta, N\to\infty$.

\end{document}